\documentclass[usenatbib,preprint]{mn2e}
\usepackage{graphicx}
\usepackage {natbib}
\usepackage {bm}
\usepackage {amsmath,amssymb}
\usepackage {setspace}

\title{Classification of the Circumstellar Disk Evolution During the Main Accretion Phase}

\author[Tsukamoto \& Machida]{Yusuke Tsukamoto$^{1,2}$ and Masahiro N. Machida$^{2,3}$ \\
$^1$Department of Astronomy, the University of Tokyo, Hongo 7-3-1, Bunkyo-ku, Tokyo, Japan\\
$^2$Division of Theoretical Astronomy, National Astronomical Observatory of Japan 
2-21-1 Osawa, Mitaka, Tokyo, Japan\\
$^3$Department of Earth and Planetary Sciences, Kyushu University, 
6-10-1 Hakozaki, Higashi-ku, Fukuoka, Fukuoka, Japan
}

\begin{document}
\bibliographystyle{mn2e}
\maketitle

\begin{abstract}
We have carried out hydro-dynamical simulations to investigate the formation and evolution of protostar and 
circumstellar disks from the prestellar cloud.
As the initial state, we adopt the molecular cloud core with two non-dimensional parameters representing the thermal and rotational energies.
With these parameters, we make 17 models and calculate the cloud evolution $\sim 10^4$ years after the protostar formation.
We find that early evolution of the star-disk system can be qualitatively classified into four modes: the massive disk, early fragmentation, late fragmentation, and protostar dominant modes. 
In the 'massive disk mode' to which the majority of models belong, the disk mass is greater than the protostellar mass for over $10^4$ years and no fragmentation occurs in the circumstellar disk.
The collapsing cloud shows fragmentation {\it before} the protostar formation in the 'early fragmentation mode'.
The circumstellar disk shows fragmentation {\it after} the protostar formation in the 'late fragmentation mode', in which the secondary star substantially gains its mass from the circumstellar disk after fragmentation and it has a mass comparable to that of the
 primary star.
The protostellar mass rapidly increases and exceeds the circumstellar disk mass in the `protostar dominant mode'.
This mode appears only when the initial molecular cloud core has a considerably small rotational energy. 
Comparison of our results with observations indicates that a majority of protostar has a fairly massive disk during the main accretion phase: the circumstellar disk mass is comparable to or more massive than the protostar.
It is expected that such a massive disk promotes gas-giant formation by gravitational instability in a subsequent evolution stage.

\end{abstract}


\section{Introduction}
There are two major mechanisms for gas-giant planet formation: one is the core accretion mechanism in which a massive solid core forms first and the disk gas accretes onto the core  \citep{safronov69,goldreich_ward73,pollacketal96},
and the other is the gravitational instability (GI) mechanism in which the circumstellar disk directly 
fragments into gas-giant planets via GI \citep{cameron78}.
Recent discovery of extra-solar planets at a great distance from the central star such as HR8799b, c, d and e \citep{maroisetal08,maroisetal10} 
and GJ579b \citep{thalmannetal09} creates a new problem for the planet formation.
It is difficult to form planets in the regions far from central stars according to the core accretion mechanism, 
because massive solid core formation before the dissipation of gaseous disk seems to be difficult \citep{dodsonetal09}.

The gravitationally instability mechanism may be more plausible for the formation of these planets. 
Many studies of disk fragmentation have been done using either an analytic approach \citep{rafikov05} or numerical 
simulations \citep[e.g.,][]{ stamatellos_whitworth08, caietal08, boleyetal06, mejiaetal05, pickettetal03,meru_bate10}.
These efforts, however, seem to build a consensus that the planet formation by GI within $\sim 50$ AU is highly difficult 
when the disk-to-stellar mass ratio is $M_{\rm disk}/M_{\rm star}\lesssim 0.1$
whose ratio is suggested by observations  \citep[see, e.g.,][]{kitamuraetal02}.

On the other hand, \citet{imm10} showed that the circumstellar disk is comparable to or more massive than the protostar ($M_{\rm disk}/M_{\rm star}\gtrsim 1$) during the (early) main accretion phase (i.e., Class 0 or Class I stages) and is highly gravitationally unstable.
Recently, such massive disks were  observed around very young protostars \citep[e.g.,][]{eisner_carpenter06,enoch09}.
Fragmentation in such massive disks may potentially account for the formation of 
gas-giant planets with a wide separation such as HR 8799 or GJ579.
Thus, we may have to focus on fragmentation and gas-giant planet formation in a massive disk during the main accretion phase.
There are several unsolved problems: how long such a massive disk exists, or whether the massive disk can fragment to form sub-stellar objects.

So far, a few authors have investigated the early evolution of the
circumstellar disk and its fragmentation from prestellar core stage. 
Using the thin disk approximation, \citet{vorobyov_basu10a} calculated the formation and evolution of 
the circumstellar disk in the molecular cloud core  with a barotropic equation of state 
\citep[or including cooling and heating effects, ][]{vorobyov_basu10b}.
They found that a gravitationally unstable disk frequently appears in the star 
formation process and tends to show fragmentation during the main accretion phase.
With isothermal equation of state, \citet{ketal10} investigated the formation of the circumstellar disk in 
their three dimensional calculation.
They also showed massive disk formation during the main accretion phase and concluded  that the disk 
fragments if the disk-to-stellar mass ratio is greater than $M_{\rm disk}/M_{\rm star} \gtrsim 1$.
Using an adiabatic equation of state and radiative cooling, \citet{wetal09} studied the 
circumstellar disk formation with a somewhat coarse spatial resolution.
They showed that a massive disk becomes hot during the early stage of the main accretion 
phase and fragmentation is suppressed.

\citet{machidaetal10} also investigated the disk formation with a finer spatial resolution than \citet{wetal09} using a barotropic equation of state.
In the collapsing cloud, the adiabatic heating dominates the radiative cooling at $n\sim10^{10}$ cm$^{-3}$ and the first (adiabatic) core with a size of $\sim$1\,AU forms before the protostar formation \citep{larson69, mi00}.
\citet{machidaetal10} showed that the first core  formed before the protostar formation directly evolves into the circumstellar disk after the protostar formation.
Thus, to investigate the (early) formation of the disk, we need to resolve the first core that has a size of $\sim 1$\,AU.
In their calculation, however, they implicitly assumed the point-symmetric structure  because they fixed the protostellar location to the center of the computational domain.
\citet{michael_durisen10} pointed out that the stellar motion weakens GI activity in some degree, when the disk-to-stellar ratio is $M_{\rm disk}/M_{\rm star}\sim0.1$.
Thus, a more adequate treatment may be necessary for the central protostar.

All previous studies of circumstellar disk formation have shown that
the circumstellar disk can become more massive than a protostar during the mass accretion phase.
However, in such studies,  the disk evolution was investigated in a limited parameter range \citep{wetal09, machidaetal10,machida11}.
Thus, we cannot conclude whether such a massive disk that is a favorable site for gas-giant planet formation generally appears in the star formation process.
To determine this, we need to investigate the disk formation resolving $\sim$1AU structure in a wider parameter range.

The initial cloud for star formation is conventionally characterized 
by two parameters: the ratio of thermal $\alpha$ and rotational $\beta$ energy toward the gravitational energy (for detailed description, see \S 2.2).
With these parameters, many studies have calculated and classified the cloud evolution in the isothermal  \citep{boss93,miyama84,tsuribe02} and adiabatic \citep{cha_whitworth03, matsumoto_hanawa03} gas contraction  phases to investigate fragmentation {\it before} the protostar formation.
However, the evolution and fragmentation of the circumstellar disk that is 
formed {\it after} the protostar formation has not been investigated with these parameters.

In this study, with parameters of $\alpha$ and $\beta$, we simulate the formation and evolution of 
circumstellar disk from molecular 
cloud core until $\sim 10^4 $ years after the protostar formation  using a smoothed particle hydrodynamics code with sufficient spatial resolution. 
Parameters $\alpha$ and $\beta$ significantly influence the disk formation, 
because the accretion rate onto the circumstellar disk and the size and mass of the 
circumstellar disk are determined by these parameters (see, \S \ref{discussion}).
We calculate the cloud evolution for 17 models using the barotropic equation of state.

In \S \ref{method}, we describe the numerical method and initial conditions. 
The evolution of circumstellar disk with is presented in \S \ref{results}.  We discuss our results and compare them with previous works in \S \ref{discussion}. Finally, we summarize our results in \S \ref{summary}.

\section{Numerical Method and Initial Condition}
\label{method}
Our simulations are carried out using an SPH code newly developed for
this work. The code includes an individual time-steps technique and uses Barnes-Hut
tree algorithm to calculate self-gravity with opening angle $\theta=0.5$. 
We use adaptive softening length according to \citet{pm07}.
Artificial viscosity is included according to the prescription by \citet{monaghan97} with $\alpha_v=1$ plus the Balsara switch \citep{balsara95}.
Our code is parallelized with MPI.

To mimic the thermal evolution of molecular cloud, we use the barotropic equation of state as
\begin{eqnarray}
\label{eos}
P=c_{s,0}^2 \rho \left[1+\left(\frac{\rho}{\rho_c}\right)^{2/5} \right],
\end{eqnarray}
where $c_{s,0} = 190 {\rm m~ s^{-1}}$ and $\rho_c = 4 \times 10^{-14} {\rm g~ cm^{-3}}$ is adopted.
As the initial state, we take a uniform density sphere with axisymmetric density perturbation.
The density perturbation $\delta \rho$ is given as $\delta \rho=0.01\times \cos2\phi$.
The initial cloud is rigidly rotating. 
We parameterized the ratio of thermal and rotational energy to the gravitational energy  of the initial cloud.
They are characterized by two non-dimensional parameters
\begin{eqnarray}
\alpha&=&\frac{E_{\rm thermal}}{|E_{\rm grav}|}=\frac{5R_0c_{s,0}^2}{2GM},
\label{eq:alpha}
\end{eqnarray}
and
\begin{eqnarray}
\beta &=&\frac{E_{\rm rotation}}{|E_{\rm grav}|}=\frac{\Omega_0^2R_0^3}{3GM},
\label{eq:beta}
\end{eqnarray}
where $R_0$, $\Omega_0$ and $M$ are the initial cloud radius, angular velocity and mass of cloud core, respectively.
We vary parameters $\alpha$ and $\beta$ in the range of $0.4 \leq \alpha \leq 0.8$ and 
$3 \times 10^{-4} \leq \beta \leq 7 \times 10^{-2}$.
In this study, total cloud mass is fixed to $1M_\odot$ in all models to compare the evolution of cloud with the same mass.
In addition, we fixed the initial cloud temperature to 10\,K.
Thus, parameter $\alpha$ and initial cloud density are determined when the initial cloud radius is given.
Note that such treatment changes the gravitational energy of the cloud, 
After the cloud radius is fixed with arbitrary $\alpha$, parameter 
$\beta$ is determined when the initial angular velocity $\Omega_0$ is given.
The model name, parameters $\alpha$  and $\beta$, initial cloud radius,
 initial angular velocity 
and initial cloud density are listed in Table 1. 
The initial cloud is modeled with about 500 000 SPH particles.

To calculate the disk formation several $10^3 - 10^4 $ years after the protostar formation, we adopt a sink particle technique
 according to the prescription by \citet{betal95}.
Starting from the prestellar core stage, the cloud evolution is calculated without sink particle. 
Then, we assume the protostar formation and dynamically introduce a sink particle when the gas particle 
density exceeds the threshold density, $\rho_{\rm sink}=4\times 10^{-9} {\rm g~ cm^{-3}}$.
The threshold density roughly corresponds to the density at which the second collapse begins.
As shown in \citet{larson69} and \citet{mi00}, the second collapse begins when  the gas density reaches $\rho \sim 10^{-9 }{\rm g~ cm^{-3}} \sim \rho_{\rm sink}$ at which the gas temperature exceeds $T \gtrsim 2 \times 10^3$ K and molecular hydrogen begins to dissociate.
The protostar forms  immediately after the second collapse. 
Thus, in this paper, we safely define the protostar formation epoch as that at which 
the gas density exceeds  $\rho > \rho_{\rm sink}$.

To treat the gas accretion onto the sink particle after the creation of the sink particle, we set the accretion radius $r_{\rm acc}=0.85 {\rm AU}$.
Then, within the accretion radius, we allow the gas accretion onto the sink particle when the following condition is fulfilled:
(1) the gas particle density exceeds the accretion density $\rho_{\rm acc}=4 \times 10^{-11}  {\rm g~ cm^{-3}}$,
(2) it is gravitationally bound to the sink particle
 and (3)  its specific angular momentum is less than that required for it to form a circular orbit at $r_{\rm acc}$.
We did not implement a boundary condition for sink particle.

\begin{table*}
\label{initial_conditions}

\begin{center}
\caption{Model parameters and Calculation Results.}		
\begin{tabular}{ccccccccc}
\hline\hline
 Model & $\alpha$ & $\beta$ & {\it R} (AU) & $\Omega_0 ({\rm s^{-1}})$  & $\rho_{init} ~({\rm g~cm^{-3}})$  & Mode$^a$ & number of star & 
\begin{tabular}{c}
length of \\massive disk era$^b$ \\(years)
\end{tabular} \\
\hline
 1  & 0.8 & $1\times 10^{-2}$ & 7866 & $4.94 \times 10^{-14}$ & $2.9\times 10^{-19}$ & M & 1 & $ \gtrsim 1.3\times 10^{4}$ \\
 2  & 0.8 & $7\times 10^{-3}$ & 7866 & $4.14 \times 10^{-14}$ & $2.9\times 10^{-19}$ & M & 1 & $ \gtrsim 3.9\times 10^{4}$ \\
 3  & 0.8 & $3\times 10^{-3}$ & 7866 & $2.71 \times 10^{-14}$ & $2.9\times 10^{-19}$ & M & 1 & $ 1.6\times 10^{4}$ \\
 4  & 0.8 & $1\times 10^{-3}$ & 7866 & $1.56 \times 10^{-14}$ & $2.9\times 10^{-19}$ & P & 1 & $ 4.5\times 10^{3}$\\
 5  & 0.8 & $3\times 10^{-4}$ & 7866 & $8.56 \times 10^{-15}$ & $2.9\times 10^{-19}$ & P & 1 & $ 1.2\times 10^{3}$\\
 6  & 0.6 & $7\times 10^{-2}$ & 5900 & $2.01 \times 10^{-13}$ & $6.9\times 10^{-19}$ & M & 1 & $ \gtrsim 3.4\times 10^{4}$ \\
 7  & 0.6 & $1\times 10^{-2}$ & 5900 & $7.61 \times 10^{-14}$ & $6.9\times 10^{-19}$ & M & 1 & $ \gtrsim 2.0\times 10^{4}$ \\
 8  & 0.6 & $7\times 10^{-3}$ & 5900 & $6.37 \times 10^{-14}$ & $6.9\times 10^{-19}$ & M & 1 & $ \gtrsim 3.4\times 10^{4}$ \\
 9  & 0.6 & $3\times 10^{-3}$ & 5900 & $4.17 \times 10^{-14}$ & $6.9\times 10^{-19}$ & M & 1 & $ \gtrsim 1.2\times 10^{4}$ \\
 10 & 0.6 & $1\times 10^{-3}$ & 5900 & $2.41 \times 10^{-14}$ & $6.9\times 10^{-19}$ & P & 1 & $ 4.5\times 10^{3}$ \\
 11 & 0.4 & $7\times 10^{-2}$ & 3933 & $3.70 \times 10^{-13}$ & $2.3\times 10^{-18}$ & EF & 4 & \\
 12 & 0.4 & $1\times 10^{-2}$ & 3933 & $1.40 \times 10^{-13}$ & $2.3\times 10^{-18}$ & EF & 4 & \\
 13 & 0.4 & $9\times 10^{-3}$ & 3933 & $1.32 \times 10^{-13}$ & $2.3\times 10^{-18}$ & LF & 2 & \\
 14 & 0.4 & $7\times 10^{-3}$ & 3933 & $1.17 \times 10^{-13}$ & $2.3\times 10^{-18}$ & LF & 2 & \\
 15 & 0.4 & $3\times 10^{-3}$ & 3933 & $7.66 \times 10^{-14}$ & $2.3\times 10^{-18}$ & M & 1 & $ \gtrsim 1.0\times 10^{4}$ \\
 16 & 0.4 & $1\times 10^{-3}$ & 3933 & $4.42 \times 10^{-14}$ & $2.3\times 10^{-18}$ & P & 1 & $ 2.2\times 10^{3}$ \\
 17 & 0.4 & $3\times 10^{-4}$ & 3933 & $2.42 \times 10^{-14}$ & $2.3\times 10^{-18}$ & P & 1 & $ 9.0\times 10^{2}$ \\
\hline
\end{tabular}
\end{center}
\footnotesize
\begin{flushleft}
Notes:\\
$^a$ "M", "EF", "LF", "P" mean "Massive disk mode", "Early fragmentation mode", "Late fragmentation mode" and "Protostar dominant mode", respectively.\\
$^b$ "Massive disk era" is defined as the period during which the disk mass is greater than the protostar mass.
Some of simulations are terminated during the massive disk era due to computational limits and we represent it with $\gtrsim$. 
\end{flushleft}
\end{table*}

\section{Results}
\label{results}
\begin{figure*}
\includegraphics[width=130mm]{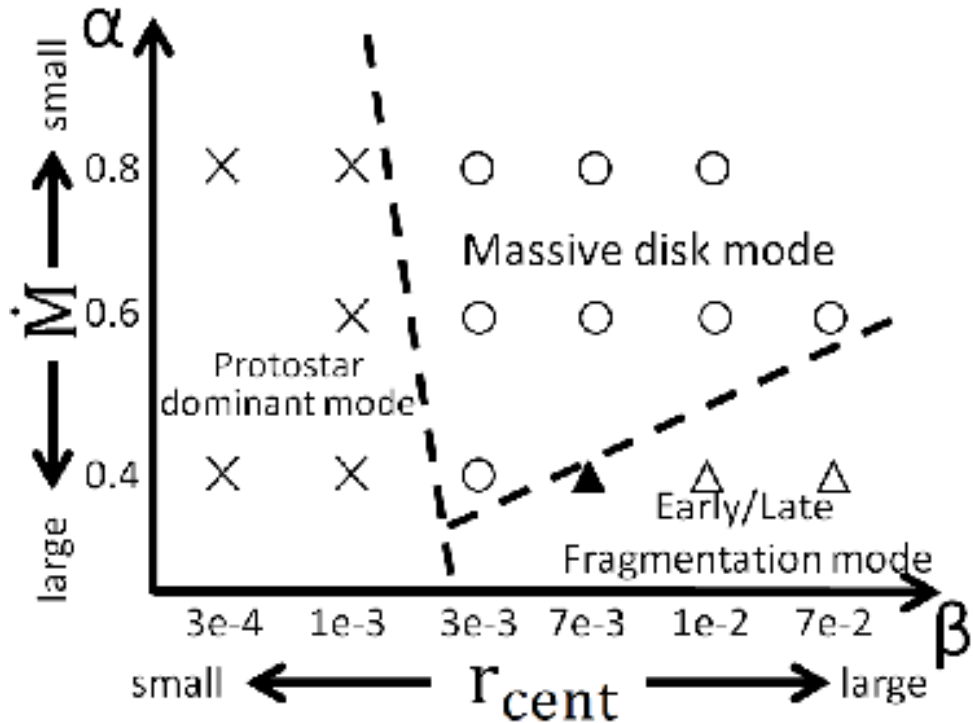}

\caption{The classification of simulation results on $\alpha$-$\beta$ plane. 
The circle, open triangle, filled triangle, cross indicate massive disk mode, early fragmentation mode, 
late fragmentation mode and protostar dominant mode, respectively. 
${\dot M}$ and ${r_{cent}}$ are the mass accretion rate
and centrifugal radius, respectively.}
\label{alpha_beta}
\end{figure*}

As listed in Table~1, we calculated the cloud evolution for 17 models in total, and classified them into four modes as follows:
\begin{enumerate}
\item Massive disk mode: the disk mass dominates the protostellar mass for over $ 10^4$ years after the protostar formation.
\item Early fragmentation mode: fragmentation occurs in the collapsing cloud  {\it before} the protostar formation.
\item Late fragmentation mode: fragmentation occurs in the circumstellar disk {\it after} the protostar formation.
\item Protostar Dominant mode: the protostellar mass rapidly dominates the disk mass within $10^4$ years after the protostar formation.
\end{enumerate}
The calculation results and its classification are summarized in Figure \ref{alpha_beta}.

\subsection{Massive Disk Mode}
\label{sec:massive}
\begin{figure*}
\includegraphics[width=130mm]{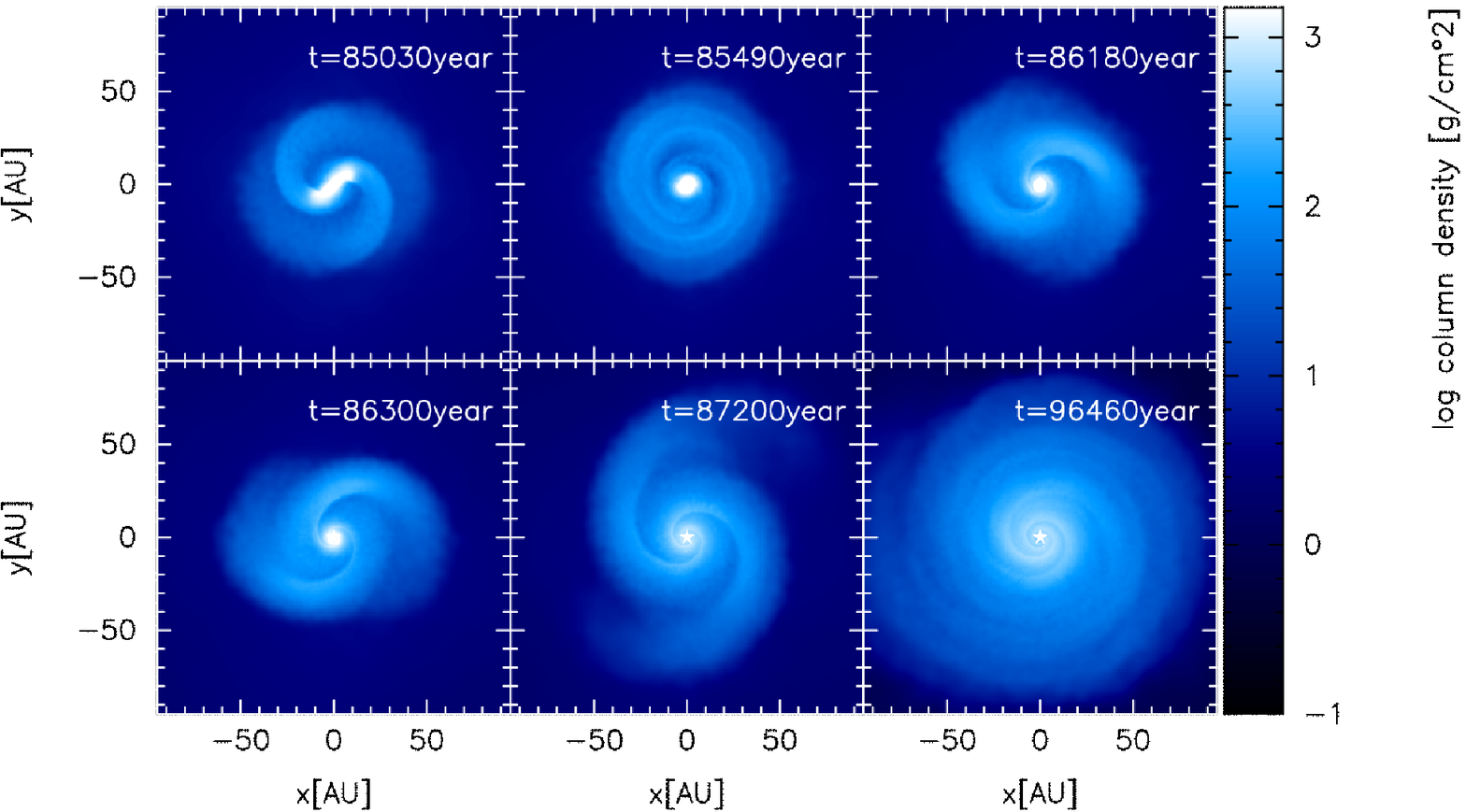}

\caption{Time sequence of the logarithm of the face-on surface density before and after the protostar formation for model 8 ($ \alpha =0.6$ and $\beta= 7 \times 10^{-3} $). 
Top left and top middle panels show the snapshot about $1.1\times 10^3$ and $6.9 \times 10^2$ years before the protostar formation. Top right panel shows the snapshot just when the protostar forms. The bottom left, middle and right show the snapshots  $1.4\times 10^2,~1.0\times 10^3$ and $1.0\times 10^4$ years after the protostar formation. The elapsed time from beginning is described in each panel. }
\label{faceon_massive_disk}
\end{figure*}

\begin{figure*}
\includegraphics[width=130mm]{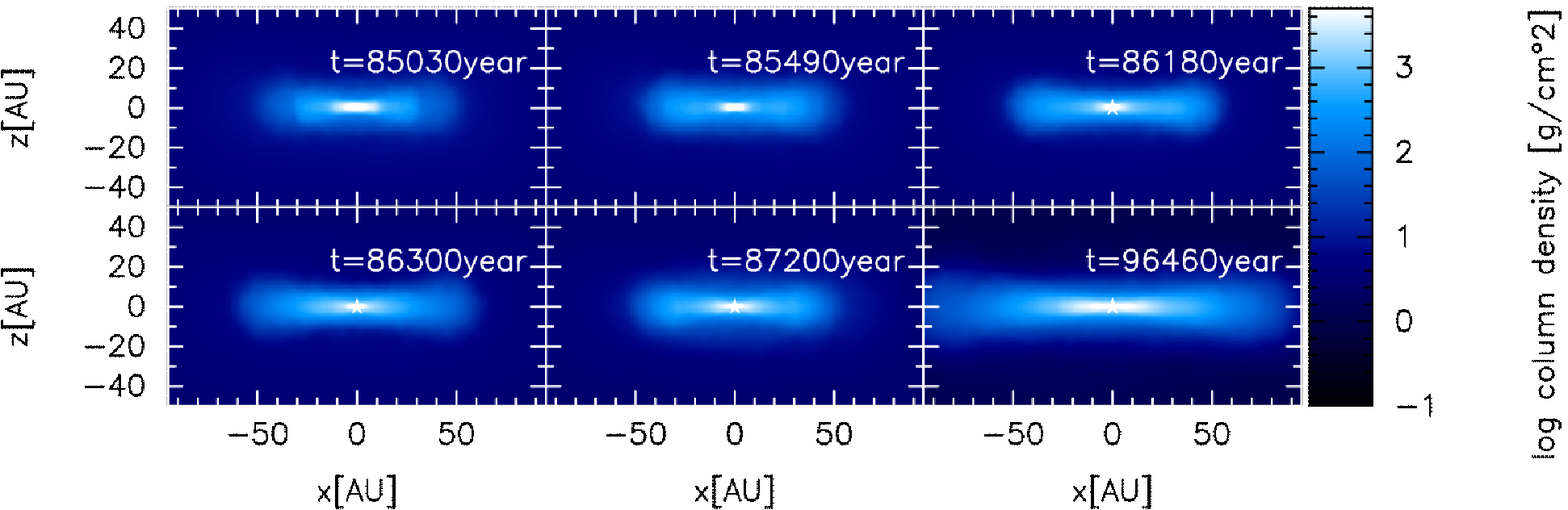}

\caption{Same as Fig.~\ref{faceon_massive_disk} but edge-on view. }
\label{edgeon_massive_disk}
\end{figure*}

Massive disk mode is indicated by the circles in Figure \ref{alpha_beta}.  
Figures \ref{faceon_massive_disk} and  \ref{edgeon_massive_disk} 
show the time evolution of the center of the cloud
for model 8 ($\alpha =0.6$ and $\beta=7 \times 10^{-3}$) that is a typical model for massive disk mode.
The upper left and upper middle panels in these figures show snapshots before the protostar formation. 
A strong bar-like structure appears in the central high-density gas region  in the top left panel in Figure~\ref{faceon_massive_disk}.
Then, the bar structure effectively transfers the angular momentum outward and the high-density gas region shrinks.
As a result, a bimodal structure composed of the central high-density core and its surrounding disk appears as seen in the top middle panel. 
These figures clearly show the disk formation before the protostar formation.
The top right panel shows the snapshot just when the second collapse occurs (i.e., the protostar forms). 
The time when the protostar forms is $8.6 \times 10^5$ years from beginning. This roughly corresponds to the free-fall timescale of the initial cloud.
By this epoch, the spiral arm has developed around the central object. 
The bottom left, middle, and right panels show the snapshots $1.4\times 10^2,~1.0\times 10^3$ and $1.0\times 10^4$ years after the protostar formation.
These figures show that the disk gradually increases its size with time retaining the global spiral structure.

\begin{figure}
\includegraphics[width=60mm,angle=-90]{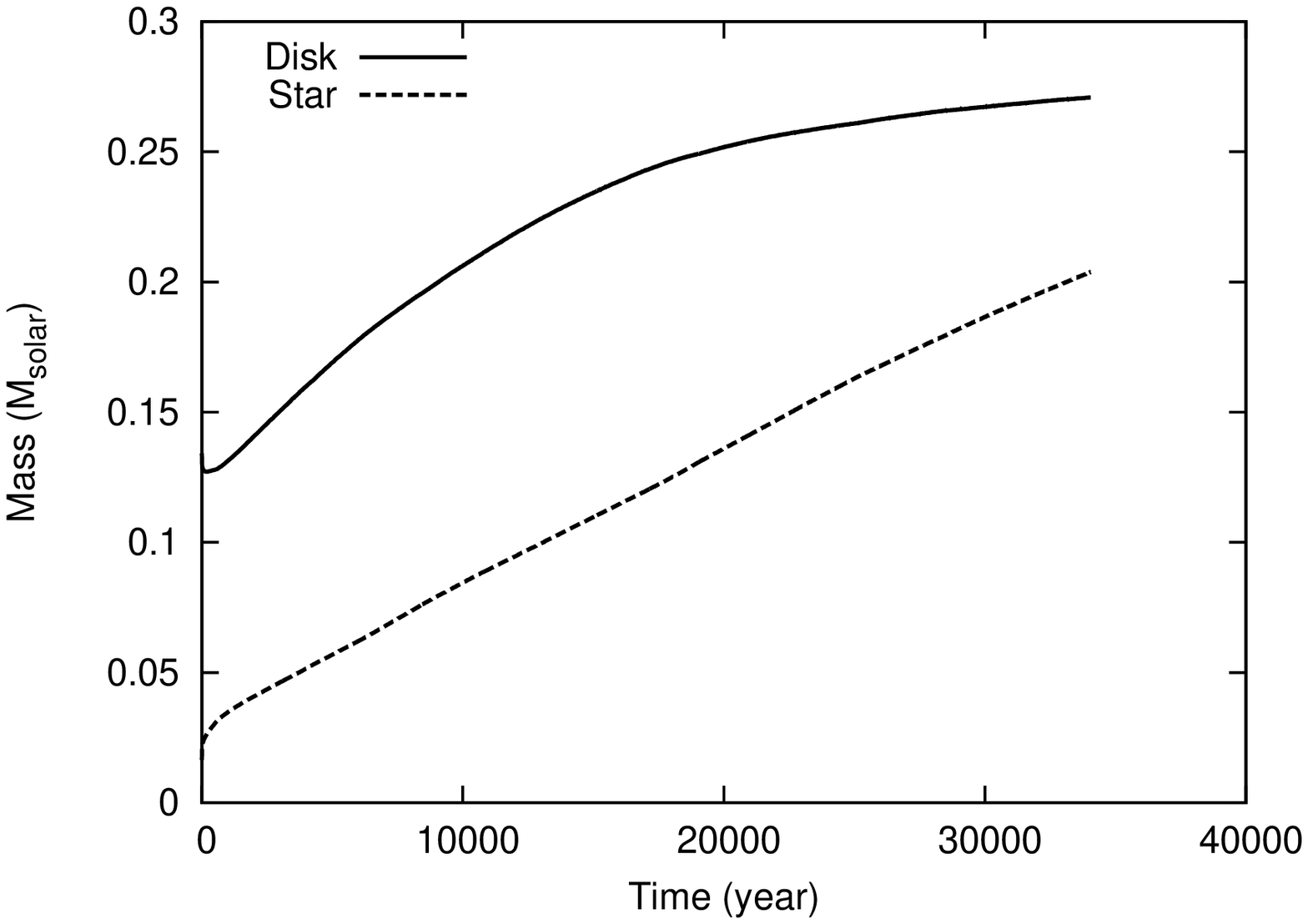}
\caption{Mass evolution of the protostar (dashed line) and the disk (solid line) for model 8. }
\label{time_mass}
\end{figure}

\begin{figure*}
\includegraphics[width=60mm]{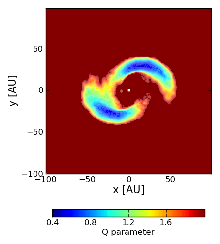}
\includegraphics[width=60mm]{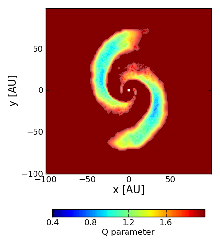}
\caption{The contours of Toomre's Q parameter at $1.4 \times 10^2$ years (left) and $1.0 \times 10^3$ years (right) after the protostar formation.
}
\label{q_value}
\end{figure*}

Figure \ref{time_mass} shows the mass evolution of the disk and protostar. For this model, the disk mass is greater than the protostellar mass for more than $10^4$ years after the protostar formation. Such a disk is expected to be gravitationally unstable.
To investigate the disk stability, the contours of Toomre's Q parameter at $1.4\times 10^2$ and $1.0\times 10^3$ years after the protostar
 formation (the same epoch of the bottom left and  middle panels of Fig.~\ref{faceon_massive_disk}) are plotted in Figure \ref{q_value}.
The Toomre's Q parameter is described as 
\begin{equation}
Q=\frac{c_s\kappa}{\pi G\Sigma},
\label{eq:Qvalue}
\end{equation}
where $c_s$, $\kappa$ and $\Sigma$ are the sound velocity and epicyclic frequency and surface density of the disk, respectively.
For this model, the circumstellar disk did not show fragmentation even though the circumstellar disk has the region of $Q<1$ at $1.4 \times 10^2$ years after the protostar formation. 
This is because GI cannot grow sufficiently fast in the circumstellar disk.
The characteristic timescale of Toomre's analysis,
$ \tau=2c_s/[G \Sigma (1-Q^2)^{1/2}] $,
is comparable to the orbital period of the disk (both are several thousand years).
In this case, there are non-linear stabilize mechanisms against GI.
As seen in the right panel of Figure \ref{q_value} or bottom middle panel of Figure \ref{faceon_massive_disk}, 
the spiral arm globally redistribute the mass and angular momentum in a short duration, $\sim 10^3$ years ($ \lesssim T_{orbit}$). 

\begin{figure}
\includegraphics[width=60mm,angle=-90]{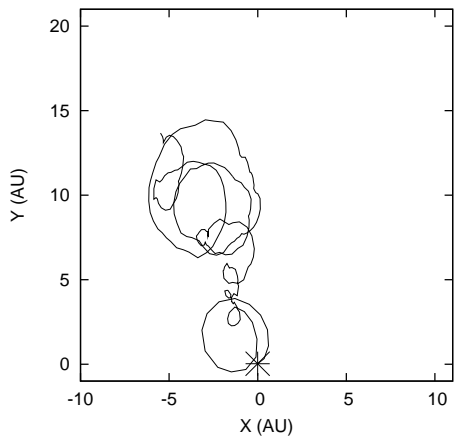}

\caption{The protostellar trajectory for $8.0 \times 10^3$ years after the protostar formation. 
The asterisk indicates the position where the protostar forms.}
\label{xy_star}
\end{figure}

Furthermore, the disk-star configuration also dynamically changes in a short duration.
Figure \ref{xy_star} shows the trajectory of the protostar for $8.0 \times 10^3$ years after the protostar formation. 
The asterisk indicates the position where the protostar forms.
The central protostar drifts toward the dense part of the spiral arm due to the gravitational interaction between 
the protostar and spiral arms. Since the strength of Keplerian shear is proportional 
to $\frac{d}{dr}(r \Omega^2) \propto r^{-3}$, its radial dependency is very strong and the drift 
motion increases the shear stress of the dense region and suppresses fragmenting of the disk.
As described in \citet{gammie01}, the thermal evolution of the disk also may play an important role 
to suppressing/promoting fragmentation.
They showed that no fragmentation occurs when the cooling time is much longer  
than the orbital period. In our simulation, since we adopted the barotropic equation of state (see, \S2.2),  the 
disk evolves adiabatically (practically all the time, the disk mid-plane density is 
greater than $\rho_c=4 \times 10^{-14}  {\rm g~ cm^{-3}}$).
With these effects, the disk is stabilized as seen in the right panel of figure \ref{q_value}. 
For disk fragmentation, the mass in-fall must be sufficiently fast to overcome these
non-linear back reactions.



\subsection{Early Fragmentation Mode}
The open triangles in Figure \ref{alpha_beta} indicate "Early fragmentation mode," in which the collapsing cloud undergoes fragmentation before the protostar formation.
Figure \ref{ring} shows the time evolution of the center of the cloud for model 11 ($\alpha =0.4$ and $\beta=7 \times 10^{-2}$).
The figure shows that the ring pattern arises (top left panel) and it fragments to form  four clumps (or protostars).
These protostars have roughly the same mass when they form. After fragmentation, they
interact with each other.
Finally, a quadruple stellar system composed of two close binary appears as seen in the lower right panel of Figure~\ref{ring}.
The condition of the ring like structure was substantially investigated in \cite{cha_whitworth03}.
Since they fixed parameter $\alpha$  (but changed the rotational low), we cannot quantitatively compare 
their results with ours. However, our result is not contradict with them.


In models showing fragmentation before the protostar formation (i.e., early fragmentation mode), fragmentation pattern qualitatively changes as the rotation energy of the initial cloud decreases.
Figure \ref{multiple} shows the time evolution of the center of the cloud for the model 12 ($\alpha =0.4$ and $\beta=1.0 \times 10^{-2}$).
A clear bar-like structure arises before the protostar formation for model 12.
The bar fragments into two clumps.
Then, a binary system appears around the center of the collapsing cloud (upper middle panel of Fig.~\ref{multiple}).
Furthermore, the remnant of the bar fragments and form two extra protostars about $10^3$ years after the first fragmentation,  as seen in the upper right panel of Figure~\ref{multiple}.
Then, one protostar is ejected from the central region by gravitational interaction among protostars, and a triple stellar system remains around the center of the cloud at the end of the calculation.

Fragmentation and subsequent evolution of fragments are very complicated in models belonging to the early fragmentation mode.
However, since fragmentation before the protostar formation was well investigated in previous studies 
(see, \citealt{tsuribe02}, \citealt{goodwin07} and \citealt{bodenheimer00} for review), we do not comment it any more in this paper.

\subsection{Late Fragmentation Mode}

\begin{figure}
\includegraphics[width=60mm,angle=-90]{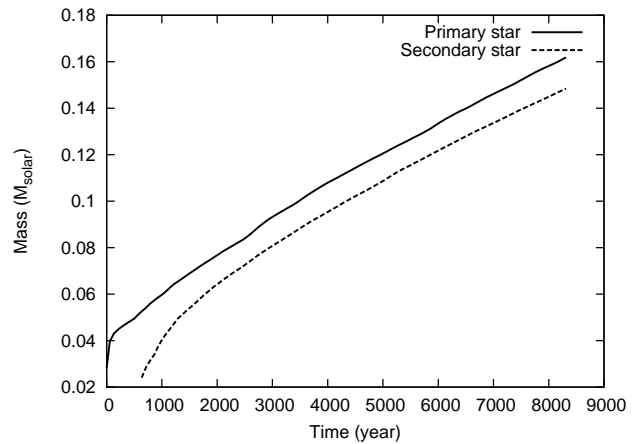}

\caption{Mass evolution of the primary star (solid line) and the secondary star (dashed line). }
\label{mass_binary}
\end{figure}

The filled triangles in Figure \ref{alpha_beta} indicates the "Late fragmentation mode," in which fragmentation occurs in the circumstellar disk after the protostar formation.
Models 13 and 14 belong to this mode.
Figure \ref{binary} shows the snapshots for model 14 ($ \alpha =0.4$ and $\beta= 7 \times 10^{-3} $).  
As seen in the massive disk mode, after a strong bar develops (top left panel),  bimodal structure composed of the central high-density gas region and disk-like structure appears even for model 14 (top middle panel).
However, fragmentation occurs in the  disk before the spiral arm sufficiently develops (top right panel) in this model, while no fragmentation occurs in the models belonging to the massive disk mode (see, Figs.~ 2 and 3).
After fragmentation, the secondary object appears in the region $\sim 30$\,AU  far from the primary star (bottom left panel)  $\sim 6 \times 10^2$ years after the primary star formation. 
Thus, Figure~\ref{binary} indicates that fragmentation can occur in the circumstellar disk during the main accretion phase if the mass in-fall is sufficiently fast (with smaller $\alpha$, see \S\ref{summary}).


The mass evolution of two protostars is plotted in Figure \ref{mass_binary}, in which the secondary star increases its mass in a way similar to that of the primary star.
At the end of the calculation, the mass of secondary star is comparable to that of the primary star.
Thus, this system evolves into a binary stellar system, not a star-planet system.

\begin{figure*}
\includegraphics[width=130mm]{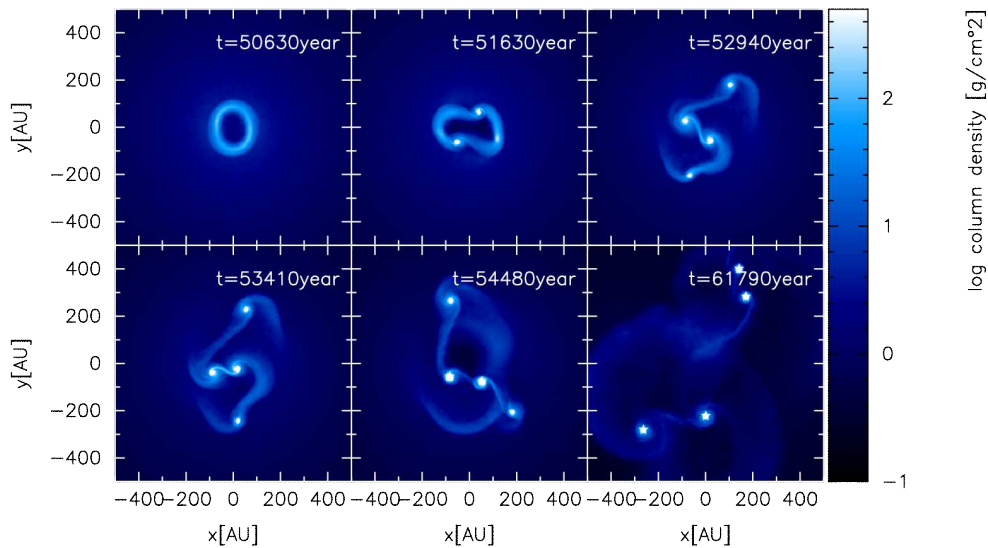}

\caption{Time sequence of the logarithm of the face-on surface density for model 11 
($ \alpha =0.4$ and $\beta= 7 \times 10^{-2} $).
The elapsed time from beginning is described in each panel.
}
\label{ring}
\end{figure*}

\begin{figure*}
\includegraphics[width=130mm]{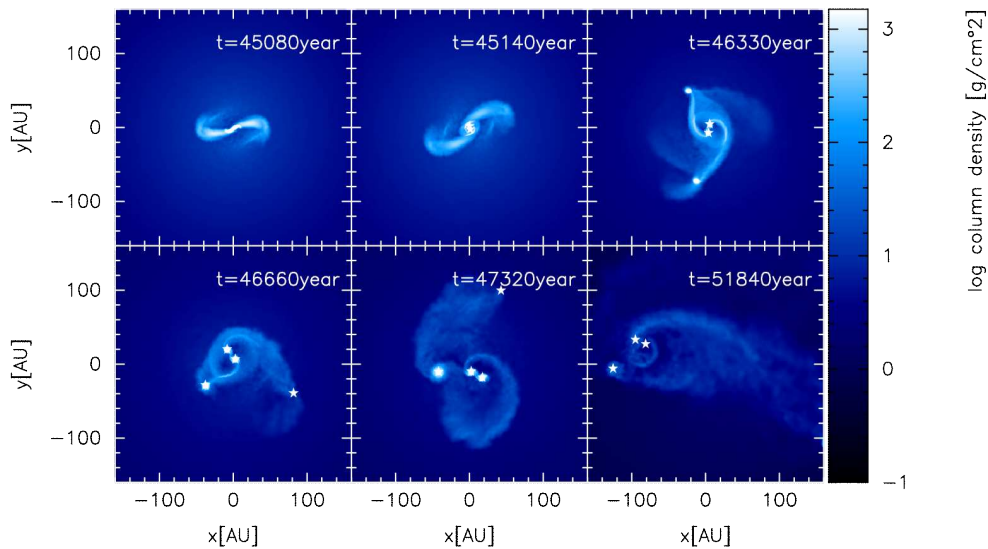}

\caption{Same as Fig.~\ref{ring} but for model 12 ($ \alpha =0.4$ and  $\beta= 1.0 \times 10^{-2} $). }
\label{multiple}
\end{figure*}

\begin{figure*}
\includegraphics[width=130mm]{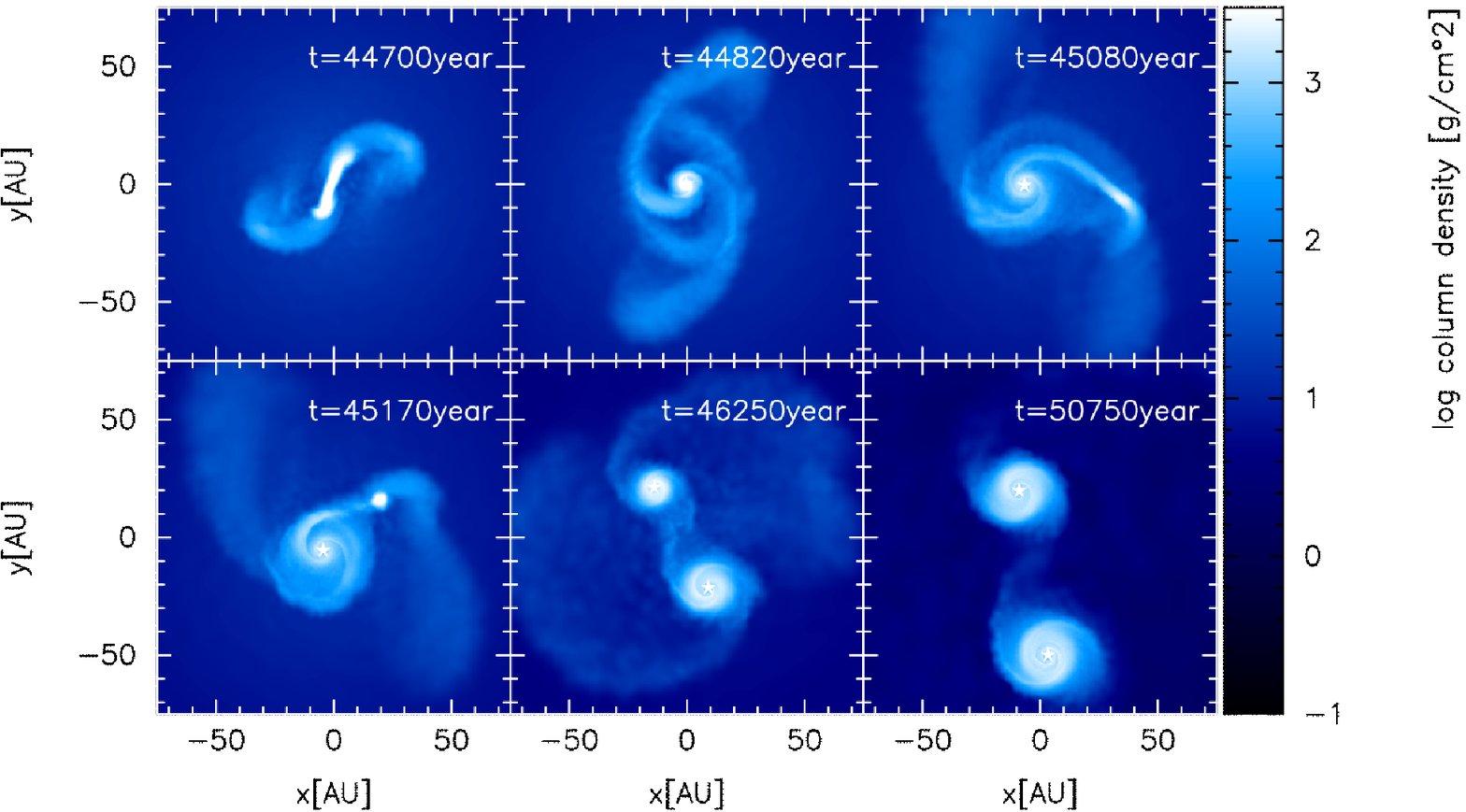}

\caption{Time sequence of the logarithm of the face-on surface density for model 14. 
Top left panel shows the snapshot about $1.2\times 10^2$ years before the protostar formation, while top middle panel shows the snapshot just when the protostar forms. The top right, bottom left, middle, 
and right show the snapshots  $2.6\times 10^2,~3.5\times 10^2,~ 1.4\times 10^3$ and $5.9\times 10^3$ years after the protostar formation. 
The elapsed time from beginning is described in each panel.
}
\label{binary}
\end{figure*}

\subsection{Protostar Dominant Mode}
\label{sec:quick}
\begin{figure*}
\includegraphics[width=130mm]{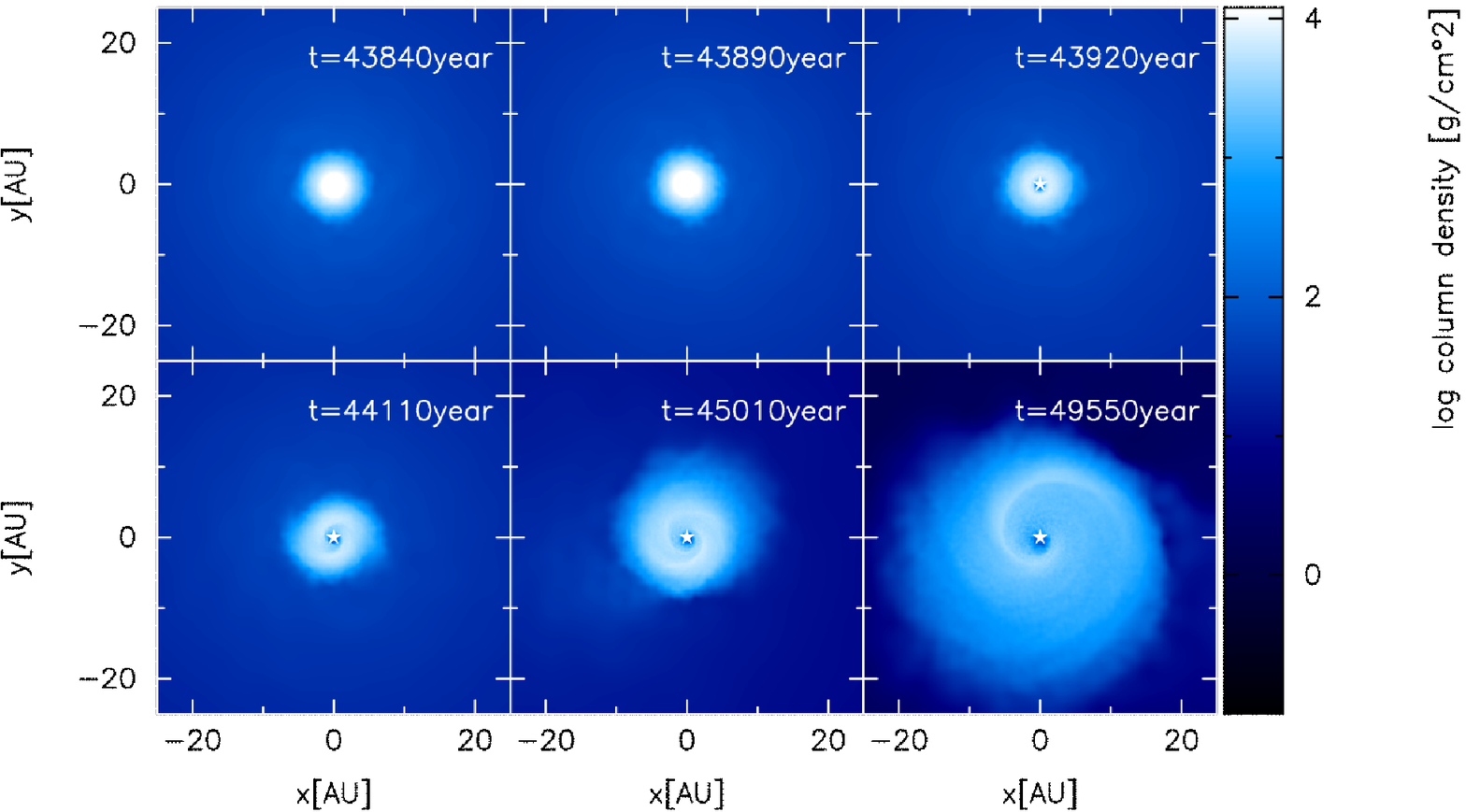}

\caption{Time sequence of the logarithm of the face-on surface density for model 16 
($ \alpha =0.4$ and $\beta= 1 \times 10^{-3} $). 
The elapsed time from beginning is described in each panel.
}
\label{faceon_classic}
\end{figure*}

\begin{figure*}
\includegraphics[width=130mm]{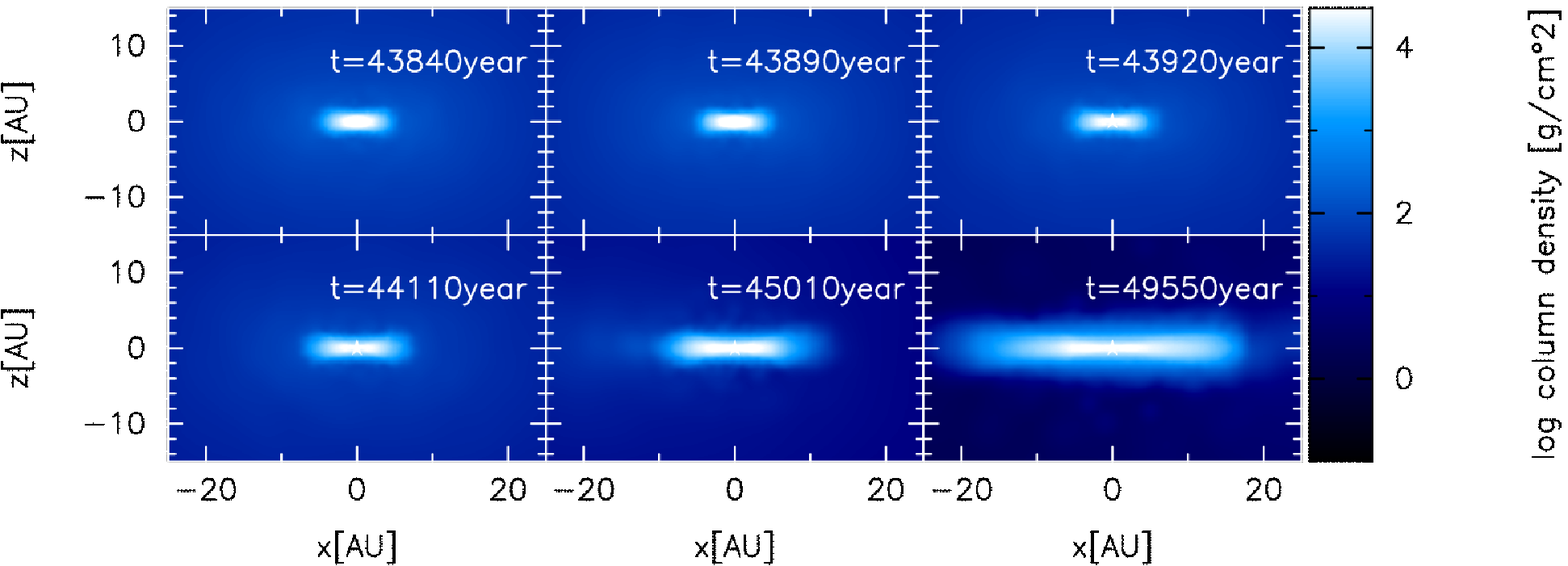}

\caption{Same as figure \ref{faceon_classic} but edge-on snapshots are shown.}
\label{edgeon_classic}
\end{figure*}

The crosses in Figure \ref{alpha_beta} indicate "protostar dominant mode", in which the protostar rapidly increases its mass and exceeds the circumstellar disk mass within $10^4$ years.
For this mode, the protostellar mass exceeds the disk mass within $10^4$ years after the protostar formation. Figures \ref{faceon_classic} and \ref{edgeon_classic} show the time evolution of the center of the cloud for model 16 
($ \alpha =0.4$ and $\beta= 10^{-3} $) that is a typical model for the protostar dominant mode.
The top left and middle panels show the snapshots before the protostar formation for model 16. 
Unlike the massive disk mode, the first core keeps an almost axisymmetric structure without 
developing the non-axisymmetric perturbation.
The second collapse occurs at $4.4\times 10^4$ years from the beginning.
Even after the protostar formation, the axisymmetric structure is maintained until the disk radius 
sufficiently grows ($r\gtrsim10$\,AU).

\begin{figure}
\includegraphics[width=60mm]{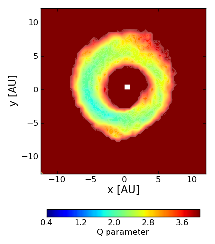}

\caption{Toomre's Q (color and contour) $1.1\times 10^3$ years after 
the protostar formation  (the same epoch as the bottom middle panel of Fig.~\ref{faceon_classic}). }
\label{q_value_classic}
\end{figure}

\begin{figure}
\includegraphics[width=60mm,angle=-90]{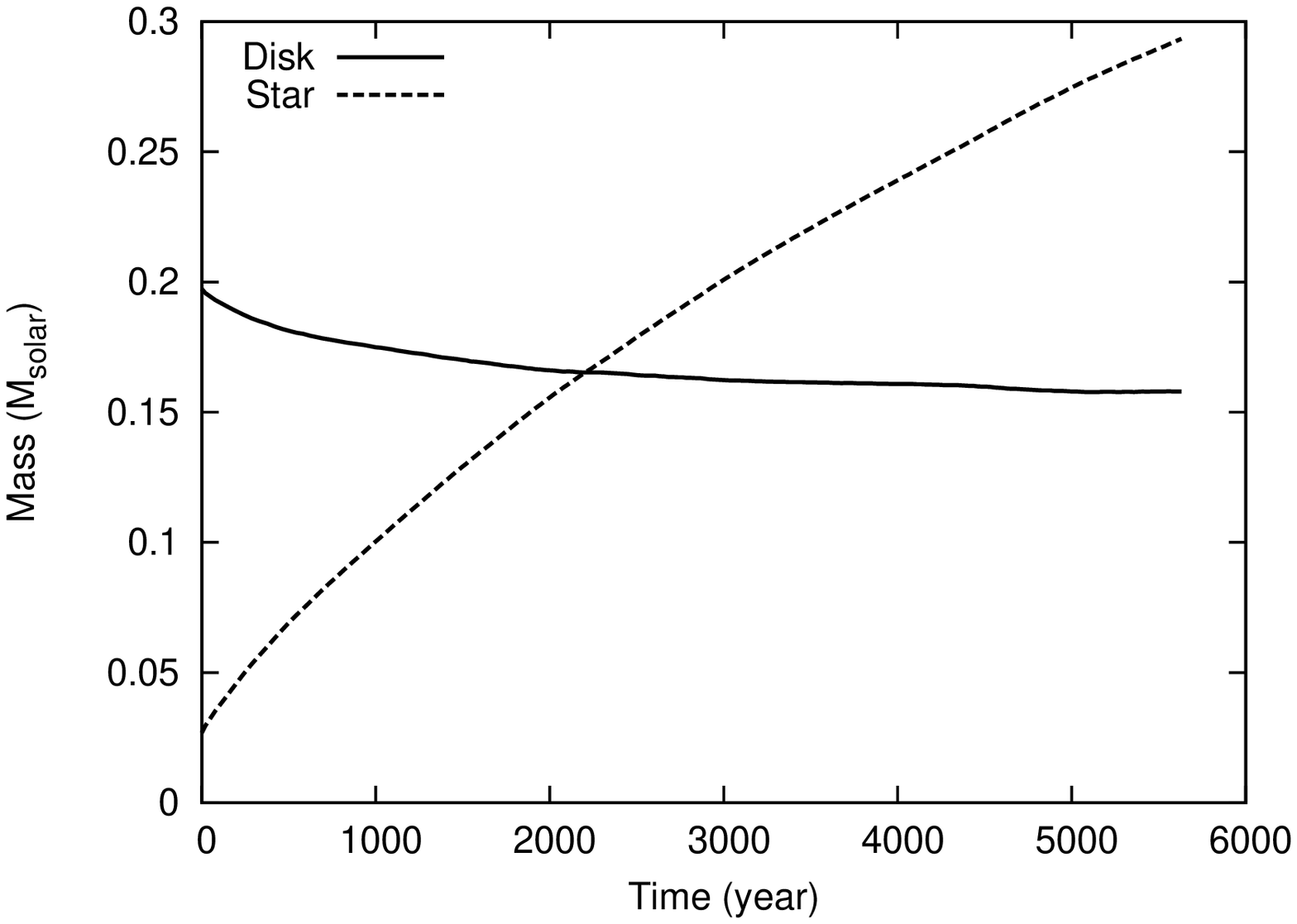}

\caption{Mass evolution of the protostar (dashed line) and the disk (solid line) for model 16.}
\label{star_disk_mass_classic}
\end{figure}

To investigate the disk stability for model 11, the Toomre's Q parameter at $1.1\times 10^3$ years after the protostar formation  
(the same epoch as the bottom middle of Fig.~\ref{faceon_classic}) is plotted in Figure~\ref{q_value_classic}. 
At this epoch, the disk mass is still greater than the protostellar mass (see, Fig. \ref{star_disk_mass_classic}). 
Nevertheless, the Q parameter is greater than unity in the whole region of the disk due to compactness of the disk.

The mass of the protostar and circumstellar disk is plotted against time after the protostar formation 
in Figure \ref{star_disk_mass_classic}.
The figure indicates that  the accretion rate onto the protostar is higher than that of the circumstellar 
disk, and the protostellar mass exceeds the circumstellar disk mass  at $\sim2.0\times10^3$ years after the protostar formation. 


\section{Discussion}
\label{discussion}

\subsection{Effect of Radiative Cooling}
\label{sec:cooling}
In this study, we used the barotropic equation of state (eq. [\ref{eos}]), in which the gas in the disk 
behaves adiabatically when the disk mid-plane density is greater than $\rho_c=4 \times 10^{-14}$\,cm$^{-3}$.
However, in reality, the disk radiatively cools with time.
According to \citet{rafikov05}, we can roughly estimate the cooling time of the disk $t_{\rm cool}$ as
\begin{eqnarray}
\label{cooling_time}
t_{\rm cool}&\simeq& \frac{\Sigma c_s^2}{\gamma - 1} \frac{f(\tau)}{2\sigma T^4} \nonumber \\
&\simeq& 
1.2\times 10^4 \left(\frac{\Sigma}{{200 \rm g~ cm^{-2}}}\right)^2 \left(\frac{T}{100 K} \right)^{-3} 
\left(\frac{\tilde{\mu}}{2.3} \right) \left(\frac{\kappa}{10{\rm cm^2~g^{-1}}} \right) ({\rm year}),
\end{eqnarray}
where $\Sigma$, $\sigma$, T, $\tau$ and $\tilde{\mu}$ are the surface density, Stephan-Boltzmann constant, mid-plane temperature, optical depth of the disk and mean molecular weight, respectively.
In equation~(\ref{cooling_time}), we estimate the vertical optical depth $f(\tau)$ of the disk as
$f(\tau) = (1 + \tau^2)/\tau.$
In addition, we assume the optical depth as
$\tau \simeq \kappa \Sigma/2$, 
where $\kappa$ is the opacity.
We assume that the disk is optically thick, and derive RHS of eq (\ref{cooling_time}) with $f(\tau)\simeq \tau$.

Equation~(\ref{cooling_time}) indicates that the cooling timescale of the circumstellar disk is typically about $10^4$ years.
Thus, the radiative cooling may play an important role in investigating the thermal evolution of the disk and its fragmentation
 over $10^4$ years.
In this study, however, we showed that the evolution of the circumstellar disk can be qualitatively classified  into four modes during very early stages of the star formation ($t\lesssim 10^4$\,years).
Thus, we expect that our classification does not change qualitatively even when the radiative effects are included.
However, the radiative cooling of disk is important to investigate the disk evolution and fragmentation for $t\gtrsim 10^4$\,years.

\subsection{Treatment of Protostar and Sink}
In this study, no model shows the formation of a star-planet system during the main accretion phase because the secondary object continues to increase its mass and finally exceeds the hydrogen-burning limit ($M\gtrsim0.08$\,M$_\odot$).
On the other hand, \citet{vorobyov_basu10a} and  \citet{machidaetal10} showed the formation of a star-planet system during the main accretion phase.
The difference is thought to be caused by treatment of the protostar and sink.

In \citet{vorobyov_basu10a} and  \citet{machidaetal10}, the protostar (or sink cell)  is fixed at the center 
of the computational domain.
It is expected that such treatment promotes fragmentation.
In reality, the density fluctuation arising around the protostar can cancel out by movement of the protostar.
Thus, fragmentation tends to occur when the protostar is fixed.

In addition, \citet{vorobyov_basu10a} and  \citet{machidaetal10} did not impose the sink on fragments formed in the circumstellar disk.
Instead, they suppressed further collapse of fragments with adiabatic equation of state.
Such treatment decreases the mass accretion onto fragments in some degree.
On the other hand, our sink treatment may overestimate the mass accretion onto fragments or protostar.

\subsection{Comparison with Previous Works}
So far, few authors have investigated the evolution of circumstellar disk from molecular cloud core. 
With an isothermal equation of state, \citet{ketal10} showed that fragmentation occurs in the circumstellar disk with a wide parameter space during the
early stage in the  main accretion phase and claimed that fragmentation frequently occurs when 
the disk-to-stellar mass ratio is greater than unity.
However, their assumption of isothermality seems not to be valid for the disk evolution of the  early main accretion phase, 
because the gas becomes opaque and behaves adiabatically when the gas density exceeds the critical density of  $\rho_c \simeq 10^{-13}-10^{- 14} {\rm g~cm^{-3}}$ (e.g., \citealt{mi00}).
In addition, the radiative cooling can affect the disk evolution $\sim 10^4$ years after the protostar formation as described in \S\ref{sec:cooling}.
Since the adiabatic equation of state  stabilizes the circumstellar disk, fragmentation barely occurs in our calculation even when the disk-to-stellar mass ratio exceeds unity.
Thus, \citet{ketal10} may overestimate the fragmentation condition, while our calculation may underestimate it  because of lack of radiative cooling. 

\citet{wetal09} also studied the circumstellar disk formation with the adiabatic 
equation of state and  radiative cooling. 
They showed no fragmentation in the circumstellar disk, because the disk becomes very hot during the early stage of main accretion phase. 
Their spatial resolution is, however, somewhat coarse; the minimum smoothing length of their study is  $h_{\rm min}= 2$ AU. On the other hand, $h_{\rm min}\sim0.3$ AU in our simulations. 
In addition, they restricted their initial conditions to rapidly rotating cases ($\beta > 10^{-2}$) to investigate 
disk evolution before the central density becomes high. 
The observation suggested that molecular cloud cores have the rotational energy of $10^{-4}<\beta< 0.07$ with a typical value of  $\beta \simeq 0.02$ \citep{cetal02}. Thus, they  studied the cloud evolution in the  limited parameter range.

\section{Summary}
\label{summary}
We have carried out hydro-dynamical simulation to investigate the evolution of the circumstellar disk with two non-dimensional parameters representing the thermal and rotational energy of the initial cloud.
The thermal energy $\alpha$ is related to the mass accretion rate onto the circumstellar disk as 
$\dot{M}_{\rm disk }= \alpha^{-3/2} c_s^3 G^{-1}$ (see, \citealt{machidaetal11}). 
Thus, smaller $\alpha$ provides a high accretion rate onto the circumstellar disk and vice versa.
On the other hand, the initial rotational energy that is represented by parameter $\beta$ is related to the disk radius. 
The centrifugal radius of the initial cloud is related to $\beta$ as
$r_{\rm cent} = 3 R_0 \beta$,
where $R_0$ is the initial cloud radius. 
Thus, with larger $\beta$, the cloud forms a larger disk in the main accretion phase. 
In other words, with larger $\beta$, a large fraction of the in-falling matter accretes 
onto the disk, rather than directly onto the primary protostar. 
As a result, smaller $\alpha$ and larger $\beta$ increase disk surface density and makes a gravitationally unstable disk.

On the other hand, the non-axisymmetric structure arose in such unstable disk can stabilize the disk, because it can redistribute the angular momentum and promote mass accretion onto the protostar.
Thus, no fragmentation occurs when a strong non-axisymmetricity grows and transfers sufficient angular momentum outward.
By contrast, the disk becomes highly gravitationally unstable and shows fragmentation  when non-axisymmetric structure does not grow sufficiently or when the growth timescale of the non-axisymmetricity is much longer than the disk growth timescale.
Thus, fragmentation condition depends also on the growth of the non-axisymmetriciy, which is closely related to parameters $\alpha$  and $\beta$ because they determine the evolution of the mass and angular momentum of the disk.

With parameters $\alpha$ and $\beta$, we found that the disk evolution is qualitatively classified into 
four modes: protostar dominant, massive disk, early fragmentation and late fragmentation modes.
The schematic classification of the circumstellar disk is shown in Figure~\ref{alpha_beta} that covers a 
wide parameter range of the rotational energy supported by observations \citep[$10^{-4}<\beta<0.07$;][]{cetal02}.
We describe each mode in the following.
\begin{itemize}
\item {\bf Protostar Dominant Mode:} 
For protostar dominant mode, the protostellar mass exceeds the circumstellar disk mass within $10^4$ years after 
protostar formation.
This mode appears in the range of $\beta\lesssim (1-3)\times 10^{-3}$ when the protostar forms, the circumstellar disk is more massive than the protostar.
However, for this mode, since the mass increase rate of the protostar is larger than that of 
the circumstellar disk, the protostellar mass exceeds the circumstellar disk mass in $\sim10^4$ 
years after the protostar formation, as described in \S\ref{sec:quick}.

\item {\bf Massive Disk Mode: }
For the massive disk mode, the circumstellar disk is more massive than the protostar
 over $10^4$ years after protostar formation.
As seen in Figure~\ref{alpha_beta}, this mode appears when the initial cloud has larger rotational and thermal energies.
In our calculation, a majority of models (8 out of 17 models) belongs to this mode.
This indicates that the circumstellar disk that is more massive than the protostar frequently appears in the star formation process.
The circumstellar disk is self-regulated for this mode.
When the circumstellar disk becomes massive, it decreases its 
mass because the non-axisymmetric structure develops and promotes the mass accretion onto the protostar.
On the other hand, when the disk is relatively less massive, the disk mass increases because a relatively 
stable disk acquires its mass from the in-falling envelope without effective angular momentum transfer.

\item {\bf Early Fragmentation Mode: }
For early fragmentation mode, fragmentation occurs before the protostar formation.
As seen in Figure~\ref{alpha_beta}, this mode appears when the initial cloud has large
rotational but small thermal energies.
If the initial rotational energy is large enough, the gas cannot condense to the center, and a ring like structure appears as seen in figure \ref{ring}. This structure is unstable against 
perturbation and it fragments into protostars.
As the rotational energy decreases, the gas can condense to the center and  bar like structure develops. 
However, unlike in the massive disk mode, since the bar cannot transfer angular momentum outward fast enough, fragmentation occurs and binary system appears as seen in Figure \ref{multiple}.

\item {\bf Late Fragmentation Mode: }
For late fragmentation mode, fragmentation occurs after the protostar formation.
The rotational energy for this mode is smaller than that for early fragmentation mode. 
Thus, without fragmentation before the protostar formation, 
the central bar like structure shrinks transferring angular momentum outward and form a single protostar and a disk.
\end{itemize}



In this study, since we did not include proper radiation treatment, we cannot discuss whether a gas-giant planet can form in such a massive disk.
However, such a massive disk is a plausible site for the gas-giant formation by GI.
To determine whether a gas-giant planet forms in the circumstellar disk by GI, we need to perform more long-term calculation including radiative effects in the future.

\section *{Acknowledgments}
We thank T. Tsuribe, S. Inutsuka, K. Osuga, K. Tomida and J. Makino for fruitful discussions.
We also thank anonymous referee for helpful comments.
The snapshots were produced by SPLASH \citep{price07}.
This work was supported by the Grants-in-Aid from MEXT (21740136).
The computations were performed on a parallel computer, XT4 system at CfCA of
NAOJ.
Y.T. is financially supported by Research Fellowships of JSPS for Young Scientists.


\begin{thebibliography}{}

\bibitem[\protect\citeauthoryear{{Bate {\it et al.}}}{{Bate {\it et al.}}}{1995}]{betal95}
Bate M.~R., Bonnell I.~A., Bromm V., 1995, MNRAS, 277, 362

\bibitem[\protect\citeauthoryear{{Balsara}}{{Balsara}}{1995}]{balsara95}
Balsara D. ~S., 1995, J. Comput. Phys., 121, 357

\bibitem[\protect\citeauthoryear{{Boss}}{{Boss}}{1993}]{boss93}
Boss A.P., 1993, ApJ., 410, 157

\bibitem[Bodenheimer et al.(2000)]{bodenheimer00} 
Bodenheimer, P., Burkert, A., Klein, R.~I., \& Boss, A.~P.\ 2000, Protostars and Planets IV, 675 

\bibitem[Boley {\it et al.} (2006)]{boleyetal06}
Boley A.C. et al, 2006, ApJ, 651, 517

\bibitem[Cai {\it et al.} (2008)]{caietal08}
Cai  K. et al, 2008, ApJ, 673, 1138

\bibitem[\protect\citeauthoryear{{Cameron}}{{Cameron}}{1978}]{cameron78}
Cameron A. G. W., 1978, M\&P, 18, 5

\bibitem[Caselli {\it et al.} (2002)]{cetal02}
Caselli P. et al, 2002, ApJ, 572, 238

\bibitem[\protect\citeauthoryear{{Cha} \& {Whitworth}}{{Cha} \&
  {Whitworth}}{2003}]{cha_whitworth03}
Cha, S. H.,,Whitworth, A. P., 2003, MNRAS, 340, 91


\bibitem[Dodson-Robinson {\it et al.} (2009)]{dodsonetal09}
Dodson-Robinson, S. E. et al, 2009, ApJ, 707, 79

\bibitem[\protect\citeauthoryear{{Eisner} \& {Carpenter}}{{Eisner} \&
  {Carpenter}}{2006}]{eisner_carpenter06}
{Eisner} J. A.,  {Carpenter} J. M., 2006, ApJ, 641, 1162

\bibitem[Enoch et al.(2009)]{enoch09} 
Enoch, M.~L., Corder, S., Dunham, M.~M., \& Duch{\^e}ne, G.\ 2009, arXiv:0910.2715 

\bibitem[\protect\citeauthoryear{{Gammie}}{{Gammie}}{2001}]{gammie01}
Gammie C. F, 2001, ApJ, 553, 174

\bibitem[\protect\citeauthoryear{{Goldreich} \& {Ward}}{{Goldreich} \&
  {Ward}}{1973}]{goldreich_ward73}
Goldreich P.,Ward, W. R.,  .,1973, ApJ, 183, 1051

\bibitem[Goodwin {\it et al.} (2007)]{goodwin07}
 Goodwin, S.~P., Kroupa, P., Goodman, A., \& Burkert, A.\ 2007, Protostars and Planets V, 133 

\bibitem[\protect\citeauthoryear{{Price} \& {Monaghan}}{{Price} \&
  {Monaghan}}{2007}]{pm07}
{Price} D.~J.,  {Monaghan} J.~J.,  2007, MNRAS, 374, 1347

\bibitem[Inutsuka {\it et al.} (2010)]{imm10}
 Inutsuka, S., Machida, M.~N., \& Matsumoto, T.\ 2010, ApJ, 718, 58

\bibitem[Kratter {\it et al.} (2010)]{ketal10}
Kratter K. M. et al, 2010, ApJ, 708, 1585


\bibitem[Kitamura {\it et al.} (2002)]{kitamuraetal02}
Kitamura Y. et al, 2002, ApJ, 581, 357

\bibitem[\protect\citeauthoryear{{Larson}}{{Larson}}{2001}]{larson69}
Larson R. B., 1969, MNRAS, 145, 271

\bibitem[\protect\citeauthoryear{{Lin} \& {Lau}}{{Lin} \&
  {Lau}}{2007}]{lin_lau79}
  {Lin} C.~C.,  {Lau} Y.~Y.,  1979, Studies in  Applied Mathematics, 60, 97

\bibitem[Machida \& Matsumoto(2011)]{machida11} 
Machida, M.~N., \& Matsumoto, T.\ 2011, MNRAS, 284 

\bibitem[Machida {\it et al.} (2010)]{machidaetal10}
Machida M. N. et al, 2010, ApJ, 724, 1006 

\bibitem[Machida {\it et al.} (2011)]{machidaetal11}
Machida M. N. et al, 2011, ApJ, 729, 42


\bibitem[Marois {\it et al.} (2008)]{maroisetal08}
Marois et al. 2008, Science, 322, 1348

\bibitem[Marois {\it et al.} (2010)]{maroisetal10}
Marois et al. 2010, Natur, 468, 1080

\bibitem[\protect\citeauthoryear{{Masunaga} \& {Inutsuka}}{{Masunaga} \&
  {Inutsuka}}{2010}]{mi00}
{Masunaga} H.,  {Inutsuka} S., 2000, ApJ, 531, 350

\bibitem[\protect\citeauthoryear{{Matsumoto} \& {Hanawa}}{{Matsumoto} \&
  {Hanawa}}{2003}]{matsumoto_hanawa03}
{Matsumoto} T.,  {Hanawa} T., 2003, ApJ, 595, 913

\bibitem[Mejia {\it et al.} (2005)]{mejiaetal05}
Mejia A.C. et al, 2005, ApJ, 619, 1098

\bibitem[\protect\citeauthoryear{{Michael} \& {Durisen}}{{Michael} \&
  {Durisen}}{2010}]{michael_durisen10}
{Michael} S.,  {Durisen} R. H.,2010, MNRAS, 406, 279

\bibitem[\protect\citeauthoryear{{Miyama}}{{Miyama}}{1984}]{miyama84}
Miyama S., 1984, ApJ, 279, 621

\bibitem[\protect\citeauthoryear{{Monaghan}}{{Monaghan}}{1997}]{monaghan97}
Monaghan J,~J., 1997, J. Comput. Phys., 136, 298

\bibitem[\protect\citeauthoryear{{Meru} \& {Bate}}{{Meru} \&
  {Bate}}{2010}]{meru_bate10}
{Meru} F.,  {Bate} R.B., 2010, MNRAS, 406, 2279

\bibitem[Pickett {\it et al.} (2003)]{pickettetal03}
Pickett B. K.,2003 ApJ, 590, 1060

\bibitem[Pollack {\it et al.} (1996)]{pollacketal96}
Pollack et al. 1996, Icar, 124, 62

\bibitem[Price (2007)]{price07}
Price D. ,2007, PASA, 24, 159

\bibitem[\protect\citeauthoryear{{Rafikov}}{{Rafikov}}{2005}]{rafikov05}
Rafikov R. R., 2005, ApJ, 621, 69

\bibitem[\protect\citeauthoryear{{Safronov}}{{Safronov}}{1969}]{safronov69}
Safronov  V.~S., 1969, QB, 981, 26

\bibitem[\protect\citeauthoryear{{Saigo} \& {Tomisaka}}{{Saigo} \&
  {Tomisaka}}{2006}]{saigo_tomisaka06}
{Saigo} K.,  {Tomisaka} K., 2006, ApJ, 645, 381


\bibitem[\protect\citeauthoryear{{Stamatellos} \& {Whitworth}}{{Stamatellos} \&
  {Whitworth}}{2007}]{stamatellos_whitworth08}
{Stamatellos} D.,  {Whitworth} A.P., 2008, A\&A, 480, 879

\bibitem[{Tsuribe} (2002)]{tsuribe02}
Tsuribe T, 2002, PThPS, 147, 155

\bibitem[Thalmann {\it et al.} (2009)]{thalmannetal09}
Thalmann et al., 2009, ApJ, 707, 123

\bibitem[\protect\citeauthoryear{{Vorobyov} \& {Basu}}{{Vorobyov} \&
  {Basu}}{2010a}]{vorobyov_basu10a}
{Vorobyov} E. I.,  {Basu} S, 2010, ApJL, 714, 133

\bibitem[\protect\citeauthoryear{{Vorobyov} \& {Basu}}{{Vorobyov} \&
  {Basu}}{2010b}]{vorobyov_basu10b}
{Vorobyov} E. I.,  {Basu} S, 2010, ApJ, 719, 1896


\bibitem[Walch {\it et al.} (2009)]{wetal09}
Walch S. et al., 2010, MNRAS, 402, 2253



\end{thebibliography}
\end{document}